\documentclass[11pt]{article}
\usepackage{graphicx}
\usepackage{epsfig}
\usepackage{feynmf}
\usepackage{here}
\usepackage{wrapfig}
\usepackage{psfig}
\usepackage{psfrag}
\usepackage{lscape}
\usepackage{rotating}
\usepackage{amsmath}
\usepackage{amssymb}

\newcommand{\ls}{\ensuremath{e^\pm e^\pm\ }}
\newcommand{\uls}{\ensuremath{e^+ e^-\ }}
\def\geantf      {\mbox{\tt GEANT4}\xspace}
\newcommand{\BABARPubYear}    {04}

\newcommand{\BABARProcNumber} {022}
\newcommand{\SLACPubNumber} {10503}
\newcommand{\LANLNumber} {0406050}

\input babarsym

\def\Bxclnu {\ensuremath{B \ra X_c \ell \bar{\nu}}}
\def\Bxlnu {\ensuremath{B \ra X \ell \bar{\nu}}}

\def\mmxn {\ensuremath{\langle M_X^n \rangle}}

\def\breco {\ensuremath{B_{\rm{reco}}}}

\def\vcb {\ensuremath{|V_{cb}|}}

\long\def\inst#1{\par\nobreak\kern 4pt\nobreak
    {\it #1}\par\vskip 10pt plus 3pt minus 3pt}

\begin{document}
{\pagestyle{empty}

\begin{flushright}
SLAC-PUB-\SLACPubNumber \\
\babar-PROC-\BABARPubYear/\BABARProcNumber \\
hep-ex/\LANLNumber \\
June, 2004 \\
\end{flushright}

\par\vskip 4cm

\begin{center}
\Large \bf Inclusive Semileptonic \boldmath$B$ Decays at \babar\ and Extraction of HQE Parameters
\end{center}
\bigskip

\begin{center}
\large 
H.U. Flaecher\\
Department of Physics, Royal Holloway, University of London \\
Egham, Surrey, TW20 0EX, UK \\
(for the \lbabar\ Collaboration)
\end{center}
\bigskip \bigskip

\begin{center}
\large \bf Abstract
\end{center}
A measurement of the first four moments of the hadronic mass distribution
in \Bxclnu\ decays is presented for minimum lepton momenta varying between 0.9 and 1.6 
\gev, using data recorded with the \babar\ detector.
Furthermore, a measurement of the inclusive electron energy spectrum for semileptonic $B$ decays together with a measurement of its first, second and third moments for minimum electron energies between 0.6 and 1.5 \gev is reported.
We determine the inclusive \Bxclnu\ branching fraction, ${\cal B}_{c\ell\nu}$,
the CKM matrix element $|V_{cb}|$, and other heavy-quark parameters from a simultaneous 
fit to the measured moments. 
\vfill
\begin{center}
Contributed to the Proceedings of the XXXIX$^{th}$ Rencontres de Moriond on QCD
and High Energy Hadronic Interactions\\
3/28/2004---4/4/2004, La Thuile, Italy
\end{center}

\vspace{1.0cm}
\begin{center}
{\em Stanford Linear Accelerator Center, Stanford University, 
Stanford, CA 94309} \\ \vspace{0.1cm}\hrule\vspace{0.1cm}
Work supported in part by Department of Energy contract DE-AC03-76SF00515.
\end{center}

\section {Introduction}

Moments of inclusive distributions and rates for semileptonic and rare $B$ decays can be related via Operator Product Expansions (OPE) \cite{Chay:1990da} 
to fundamental parameters of the Standard
Model, such as the Cabibbo-Kobayashi-Maskawa matrix elements \Vcb and \Vub~\cite{ckm} and the heavy quark masses $m_b$ and $m_c$. 
These expansions in $1/m_b$ and the strong coupling constant $\alpha_s$ 
involve non-perturbative quantities that can be extracted from  moments of inclusive distributions. 
In the kinetic-mass scheme~\cite{kinetic} for example, these expansions to order ${\cal O}(1/m_b^3)$ 
contain six parameters: the running kinetic masses of the $b-$ and 
$c-$quarks, $m_b(\mu)$ and $m_c(\mu)$, and four non-perturbative parameters.
We determine these parameters from a fit to the moments of the hadronic mass and 
electron energy distributions in semileptonic $B$ decays to charm particles, 
\Bxclnu. 
These measurements are based on data recorded with the \babar\ detector~\cite{nimpaper} at the \FourS\ resonance.

\section{Measurement of Hadronic Mass Moments}

\subsection{Event Selection and Simulation}
The measurement of hadronic mass moments is based on a sample of 89 million \BB\ pairs corresponding to an integrated luminosity of 82 fb$^{-1}$.
The analysis uses \FourS\to\BB\ events in which one of the \B mesons decays to hadrons and is fully reconstructed ($B_{\rm reco}$)~\cite{Aubert:2003zw} and the semileptonic 
decay of the recoiling \Bb\ meson ($B_{\rm recoil}$) is identified by the presence of an electron or muon.
This approach allows for the determination of the momentum, charge, and flavor of the \B mesons.
Semileptonic decays are modeled by a parameterization of form factors for $\Bb\to D^{*}\ell^-\nub$~\cite{Duboscq:1996mv}, and models for   
$\Bb\to D \ell^-\nub,D^{**}\ell^-\nub$~\cite{ISGW2} and 
$\Bb\to D \pi \ell^-\nub, D^* \pi \ell^-\nub$~\cite{J.GoityandW.Roberts}.
Monte Carlo (MC) simulations of the \babar\ detector are based on \geantf~\cite{Agostinelli:2002hh}.

Semileptonic  decays 
are identified by the presence of exactly one electron or muon above a minimum cut-off energy $E_{\rm cut}$, measured in the rest frame of the $B_{\rm recoil}$ meson recoiling against the $B_{\rm reco}$. 
The hadronic system $X$ in the decay \Bxlnu\ is reconstructed from charged
tracks  and energy depositions in the calorimeter that are not associated with
the \breco\ candidate or the charged lepton.
The neutrino four-momentum $p_{\nu}$ is estimated from the
missing four-momentum  $p_{\rm miss} = p_{\FourS}-p_{\breco} -p_X-p_\ell$, 
where all momenta are measured in the laboratory frame. We impose the following criteria to ensure well reconstructed events: $E_{\rm miss} > 0.5 \gev$, $|\vec {p}_{\rm miss}| > 0.5 \gev$, and $|E_{\rm miss} - |\vec {p}_{\rm miss}|| < 0.5 \gev$.
We select 7114 signal events over a combinatorial background of 2102 events.
The mass of the hadronic system $M_X$ is determined by a
kinematic fit that imposes four-momentum conservation, the equality of
the masses of the two $B$ mesons, and constrains $p_{\nu}^2 = 0$. 
\vspace{-0.25cm}

\subsection{Extraction of Hadronic Mass Moments}
In order to extract unbiased moments \mmxn, 
we need to correct for effects that can distort the mass distributions.
We use observed linear relationships between the measured \mmxn\ and generated $\langle M_X^{n~{\rm true}} \rangle$ values from MC simulations in bins of $M_X^{n~{\rm true}}$ (see Fig.~\ref{fig:calib}) to calibrate the measurement of 
$M_X^n$ on an event-by-event basis.
Since any radiative photon is included in the measured hadron mass and our definition of $M_X$ does not include these photons, we employ PHOTOS~\cite{Barberio:1994qi}
to simulate QED radiative effects and correct for their impact (less than 5\%) on the moments as part of the calibration procedure.
To verify this procedure, we apply the calibration to the 
measured masses for individual hadronic states in simulated $\Bxclnu$  decays, 
and compare their calibrated mass moments to the true mass moments. 
The result of this test is also shown in Fig.~\ref{fig:calib} for $M_X$, indicating that the calibration reproduces the true moments over the full mass range.
Corresponding curves are obtained for $M_X^2$, $M_X^3$, and $M_X^4$.  
We observe no significant mass bias after calibration.  
The MC-based calibration procedure has also been validated on a data sample of partially reconstructed $D^{*+} \ra D^0 \pi^+$ decays~\cite{Aubert:2003zw} where the low-momentum $\pi^+$ serves as a tag and allows to select an inclusive event sample for which the true hadronic mass is known.
\vspace{-0.25cm}

\subsection{Results and Systematic Errors}
The hadronic mass moments \mmxn\ after background subtraction, calibration and correction for detection and selection efficiencies are shown in the upper half of Fig.~\ref{fig:moments}. The full numerical results can be found in~\cite{hadrmass}.
The four moments increase as $E_{\rm cut}$ decreases due
to the presence of higher mass charm states. The moment measurements are highly correlated. 
The statistical and systematic errors are of comparable size. The dominant systematic error sources stem from the precision of the modeling of the detector efficiency and particle reconstruction, the subtraction of the combinatorial background of the $B_{\rm reco}$ sample and remaining $B$ background, and uncertainties in the modeling of the hadronic states.

 \begin{figure}[t]
    \begin{centering}
	\epsfig{file=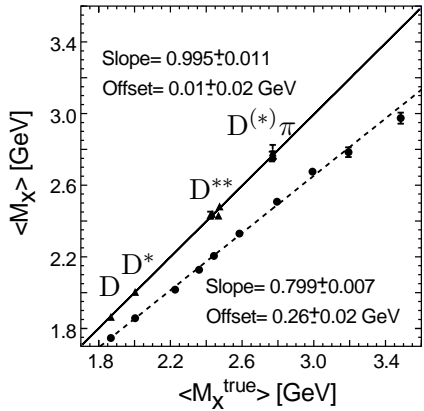,height=5.5cm}
\unitlength1.0cm
\put(-4.3,1.6){$\rm D$}
\put(-4.0,1.9){$\rm D^*$}
\put(-3.1,2.9){$\rm D^{**}$}
\put(-2.6,3.75){$\rm D^{(*)}\pi$}
     \caption{ \label{fig:calib} Results of the $\langle M_X \rangle$ calibration procedure.
The calibration data and fit results are shown by the lower dashed line (circles), the verification by the upper solid line (triangles).   }
   \end{centering}
\vspace{-0.3cm}
   \end{figure}

\section{Measurement of Electron Energy Spectrum and its Moments}

\subsection{Event Selection}

The measurement of the electron energy spectrum and its moments is based on a data sample of 47.4 fb$^{-1}$ recorded at the \FourS\ resonance and 9.1 fb$^{-1}$ at an energy 40 \mev\ below the resonance, measured in the electron-positron center of mass frame.
The analysis is similar to the \babar \ measurement of the semileptonic branching 
fraction~\cite{oldprl}, but
supersedes it by an order of magnitude in integrated luminosity.

We identify \BB\ events by observing
an electron, $e_{tag}$, with a momentum of $1.4<p^*<2.3~\gevc$ in the 
$\Upsilon(4S)$ rest frame. These electrons make up the tag-sample
which is used as normalization for the branching fraction.
A second electron, $e_{sig}$, for which we require 
$p^*>0.5 \gevc$ is assigned either to the unlike-sign sample (charge $Q(e_{tag})=-Q(e_{sig})$) 
or to the like-sign sample ($Q(e_{tag})=Q(e_{sig})$). 
In events without \BzBzb mixing, primary electrons from semileptonic 
\B decays belong to the unlike-sign sample
while secondary electrons contribute to the like-sign sample.
We select \BB events by making requirements on the charged and neutral multiplicities and event shape variables.
Electrons originating from the same \B meson as the tag electron are rejected by requiring $\cos\alpha^*   > 1.0 - p_e^*{\rm (\gev/c)}~\rm{and} \cos\alpha^*> -0.2$, where $\alpha^*$ is the opening angle between the two electrons.
Backgrounds from $\jpsi\to e^+e^-$ 
decays to the tag-sample, are suppressed by requiring
the invariant mass $M_{ee}$ of the tag electron, paired with any electron 
of opposite charge with  $\cos\alpha^*<-0.2$, to be outside the interval  
$2.9 < M_{ee} < 3.15 \gevcc$. The efficiencies of these selection criteria 
are estimated by MC simulation.

\subsection{Measurement of the Electron Momentum Spectrum}

In order to extract the electron momentum spectrum remaining backgrounds have to be subtracted.
Continuum background is subtracted from the tagged, like- and unlike-sign 
samples by scaling the 
off-resonance yields by the ratio of on- to off-resonance integrated 
luminosities, corrected for the energy dependence of the continuum 
cross section.
Background electron spectra from photon conversions and Dalitz decays are 
extracted from data.
Further backgrounds arise from decays of $\tau$ leptons, charmed 
mesons produced in $b \ra c\cbar s$ decays and $\jpsi$ or 
$\psitwos \to \epem$ decays with only one detected $e$. These backgrounds
are irreducible, and their contributions to the three electron samples are 
estimated from MC simulations.

To account for \BzBzb\ mixing, we determine the number of primary electrons in 
the $i$-th $p^*$ bin from the like-sign and unlike-sign pairs as
\begin{equation}
N^i_{b \to c,u}=\frac{1-f_0\chi_0}{1-2f_0\chi}\frac{N_{\uls}^i}
{\epsilon_{\alpha^*}^i}-\frac{f_0\chi_0}{1-2f_0\chi_0}N^i_{\ls}
\label{eq:prompt}
\end{equation} 
where $\chi_0$ is the \BzBzb\ mixing parameter 
$\chi_0 = 0.186 \pm 0.004$~\cite{PDG03} and 
$f_0=\BR(\FourS \to \BzBzb)$ = $0.490\pm 0.018$~\cite{PDG03}.
The parameter $\epsilon_{\alpha}^i$ is the efficiency of the additional
requirement on the opening angle for the unlike-sign sample.
The spectrum of primary electrons is shown in Fig.~\ref{fig_spectrum} and is corrected for bremsstrahlung in the detector using MC simulation.
\vspace{-0.1cm}

\begin{figure}[t]
\begin{center}
\epsfig{file=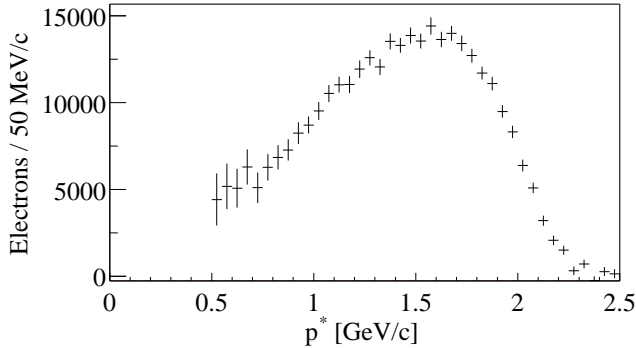, height=1.8in} 
\caption {Electron momentum spectrum from $\B \to X e \nu (\gamma)$ decays in the
$\Upsilon(4S)$ frame after correction for efficiencies and bremsstrahlung, with 
combined statistical and systematic errors. 
\label{fig_spectrum}}
\end{center}
\end{figure} 

\subsection{Extraction of Electron Energy Moments}
Before extraction of the moments as a function of the minimal 
electron energy ranging from 0.6 \gev to 1.5 \gev, 
the measured electron momentum spectrum has to be corrected for the 
contribution from $B \ra X_u e \bar{\nu}$ decays.
Defining
$R_i(E_{\rm cut},\mu)$ as $\int_{E_{\rm cut}}^{\infty} (E_e-\mu)^i (d\Gamma/dE_e)\,dE_e$,
we present measurements of the partial branching fraction $M_0(E_{\rm cut})$, the first moment $M_1(E_{\rm cut}) = R_1(E_{\rm cut},0) / R_0(E_{\rm cut},0)$ and
the central moments $M_n(E_{\rm cut})=R_n(E_{\rm cut},M_1(E_{\rm cut}))/R_0(E_{\rm cut},0)$ for $n$ = 2, 3.
Furthermore, the moments are corrected for the movement of the \B mesons in the center of mass frame and QED radiative effects. 
The results are shown in the lower part of Fig.~\ref{fig:moments}.
The full numerical results can be found in~\cite{lepton}.
The main systematic errors stem from uncertainties in the branching fractions 
of the irreducible background, electron identification efficiency and subtraction of the $B \ra X_u e \bar{\nu}$ background.
\vspace{-0.05cm}

\section{Determination of the Branching Fraction for ${\boldmath \Bxclnu}$ Decays and of ${\boldmath |V_{cb}|}$ from Hadronic Mass and Lepton Energy Moments}

\begin{figure}[t]
\begin{centering}
\unitlength 1cm
\epsfig{file=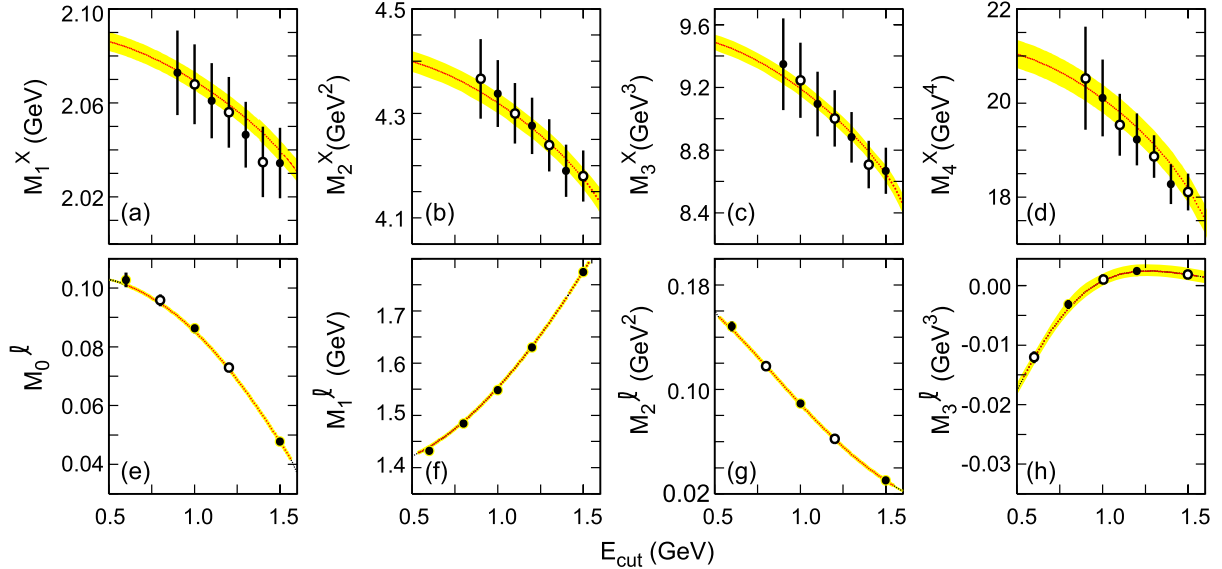, width=16cm}
\caption{\label{fig:moments} The measured hadronic-mass (a-d) and electron-energy (e-h) moments as a function of the cut-off energy, $E_{cut}$,
compared with the result of the simultaneous fit (line), with the theoretical uncertainties
indicated as shaded bands.  
The solid data points mark the measurements included in the fit.  The vertical bars indicate the experimental errors;
 $i.e.$, the statistical and systematic errors added in quadrature; 
in some cases they are comparable in size to the data points.  Moment measurements for different $E_{cut}$ are highly correlated.}
\end{centering}
\end{figure}

The $E_{\rm cut}$-dependent moment measurements discussed in the previous two sections are used to extract the total branching fraction ${\cal B}_{c \ell \nu}$, $|V_{cb}|$ and other heavy quark parameters from a simultaneous $\chi^2$ fit of OPE calculations in the kinetic mass scheme. In this scheme, the rate of \Bxclnu\ can be expressed as~\cite{kolya} 
\begin{eqnarray}
\label{equ:opegsl}
\nonumber
\Gamma_{c \ell \nu} = \frac{G_F^2 m_b^5}{192\pi^3} |V_{cb}|^2 (1+A_{ew}) A_{pert}(r,\mu)\, \times &\nonumber \\
\Bigg [ z_0(r) \Bigg ( 1 - \frac{\mu_{\pi}^2-\mu_G^2+\frac{\rho_D^3+\rho_{LS}^3}{m_b}}{2m_b^2} \Bigg ) - 2(1-r)^4 &\frac{\mu_G^2+\frac{\rho_D^3+\rho_{LS}^3}{m_b}}{m_b^2}+d(r)\frac{\rho_D^3}{m_b^3}+{\cal O}(1/m_b^4)\Bigg].
\end{eqnarray}
to ${\cal O}(1/m_b^3)$.
The leading non-perturbative effects
arise at ${\cal O}(1/m_b^2)$ and are parameterized by $\mu_{\pi}^2(\mu)$ and $\mu_{G}^2(\mu)$,  the expectation values of the kinetic and chromomagnetic dimension-five operators. 
At ${\cal O}(1/m_b^3)$, two additional parameters enter, 
$\rho_{D}^3(\mu)$ and $\rho_{LS}^3(\mu)$, the  expectation values 
of the Darwin ($D$) and spin-orbit ($LS$) dimension-six operators. 
These parameters depend on the scale $\mu$ that  separates short-distance from long-distance QCD effects; the calculations are performed for $\mu=1\gev$~\cite{kinetic}. The ratio $r=m_c^2/m_b^2$ enters in the phase-space factor $z_0(r)$ and the function $d(r)$~\footnote{Analytical expressions for $z_0(r)$ and $d(r)$ can be found in~\cite{interpret}.}.
HQEs in terms of the same heavy-quark parameters are available for the hadronic mass and electron energy moments~\cite{gambino}.
Since many of these individual moments are highly correlated we select for the fitting procedure a set of moments for which the correlation coefficients are less than 95\%.  Thus we only use half of the 28 mass moments, 
and retain 13 of the 20 energy moments.

The global fit takes into account the statistical and systematic errors and correlations of the individual measurements, as well as the 
uncertainties of the expressions for the individual moments.
The resulting fit, shown in Fig.~\ref{fig:moments}, describes the data well with $\chi^2=15.0$ for 20 degrees of freedom.  
Table~\ref{tab:results} lists the fitted parameters and their errors~\footnote{The correlations between the fit parameters can be found in~\cite{interpret}.}.
An additional error on \vcb\ has been derived from the limited knowledge of the OPE expression for the decay rate, including various perturbative corrections and higher-order non-perturbative corrections~\cite{kolya}.
As expected, $m_b$ and $m_c$ are highly correlated and
for the mass difference we obtain $ m_b - m_c=(3.436 \pm 0.025_{exp} \pm 0.018_{HQE} \pm 0.010_{\alpha_s}) \gev$. 

Several crosschecks have been carried out to ensure that the fit results are unbiased. These include variations of the theoretical uncertainties and the set of moment measurements used in the fit. In particular these have been split into sets above and below a cut-off energy of 1.2 \gev\ and hadron mass and lepton energy moments.
All results agree with each other within errors.
Figure~\ref{fig:ellipses} shows the $\Delta \chi^2=1$ ellipses for $|V_{cb}|$ versus $m_b$ and $\mu_{\pi}^2$, for a fit to all moments and separate fits to the electron energy moments and the hadronic mass moments, but including the partial branching fractions in both.

\begin{table}[t]
\caption{\label{tab:results} Fit results and error contributions from the 
moment measurements, approximations to the HQEs, and additional theoretical uncertainti
es from $\alpha_s$ terms and other perturbative and non-perturbative terms contributing
 to $\Gamma_{c\ell\nu}$.}
\footnotesize
\begin{tabular}{|ccccccccc|}
\hline
&$|V_{cb}| (10^{-3})$ 
&$m_b \mathrm{(GeV)}$ 
&$m_c \mathrm{(GeV)}$ 
&$\mu_{\pi}^2 \mathrm{(GeV^2)}$ 
&$\rho_D^3 \mathrm{(GeV^3)}$ 
&$\mu_{G}^2 \mathrm{(GeV^2)}$ 
&$\rho_{LS}^3 \mathrm{(GeV^3)}$ 
&${\cal B}_{c\ell\nu} (\%)$    \\ \hline
             Results &41.390 &4.611 &1.175  &0.447 &0.195 &0.267 &-0.085 &10.611    \\ 
\hline 
        $\delta_{exp}$ & 0.437 &0.052 &0.072 &0.035 &0.023 &0.055 &0.038 & 0.163 \\ 
        $\delta_{HQE}$ & 0.398 &0.041 &0.056 &0.038 &0.018 &0.033 &0.072 & 0.063 \\ 
   $\delta_{\alpha_s}$ & 0.150 &0.015 &0.015 &0.010 &0.004 &0.018 &0.010 & 0.000 \\ 
   $\delta_{\Gamma}  $ & 0.620 &      &      &      &      &      &      &       \\ \hline
        $\delta_{tot}$ & 0.870 &0.068 &0.092 &0.053 &0.029 &0.067 &0.082 & 0.175 \\ \hline
\end{tabular}
\end{table}

\begin{figure}[t]
\begin{center}
\epsfig{file=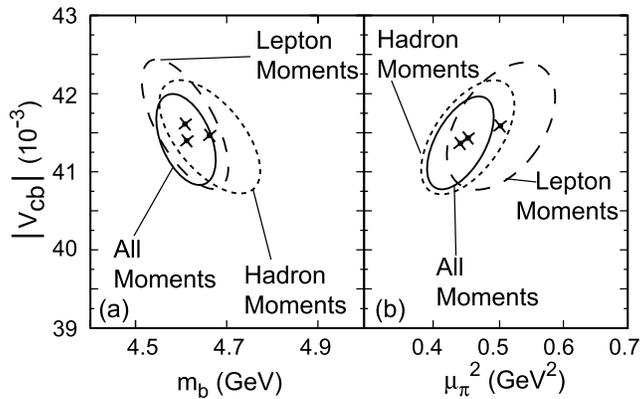, width=8.5cm}
\caption{
Fit results (crosses) with contours corresponding to $\Delta \chi^2=1$ for two pairs of
 the eight free parameters
a) $m_b$ and b) $\mu_{\pi}^2$ versus $|V_{cb}|$, separately for fits using the hadronic
-mass, the electron-energy, and all moments.
}
\label{fig:ellipses}
\end{center} 
\end{figure}

\section{Conclusions}

From the measured hadronic mass and electron energy moments we have determined the
semileptonic branching fraction, \vcb\ and the heavy quark masses $m_b$ and $m_c$.
\begin{eqnarray}
\nonumber
|V_{cb}|&=& (41.4 \pm 0.4_{exp} \pm  0.4_{HQE} \pm 0.6_{th})\,\times 10^{-3},  \nonumber \\
{\cal B}_{c e \nu} &=& ( 10.61 \pm 0.16_{exp} \pm 0.06_{HQE}) \%, \nonumber  \\
m_b(1 \gev)&=&(4.61 \pm 0.05_{exp} \pm 0.04_{HQE} \pm 0.02_{th}) \gev, \nonumber \\
m_c(1 \gev)&=&(1.18 \pm 0.07_{exp} \pm 0.06_{HQE} \pm 0.02_{th}) \gev, \nonumber
\end{eqnarray}
In addition, the non-perturbative parameters in the kinetic scheme were determined up to order $1/m_b^3$ without applying any external constraints and the results are in agreement with theoretical estimates. Consistent results were found from separate fits using hadronic mass or lepton energy moments only which gives confidence in the reliability of the OPE calculations.


\end{document}